\documentclass[preprint2]{emulateapj}
\usepackage{rotating}

\slugcomment{The Astronomical Journal}

\shorttitle{Photometric and Proper Motion Study of NGC 2215}
\shortauthors{Fitzgerald et al.}

\begin{document}


\title{Photometric and proper motion study of \\ neglected open cluster NGC 2215}


\author{M.T. Fitzgerald\altaffilmark{1,2}, L. Inwood \altaffilmark{3}, D.H. McKinnon\altaffilmark{2}, W. S. Dias\altaffilmark{5,6}, M. Sacchi\altaffilmark{4},\\ B. Scott\altaffilmark{7},M. Zolinski\altaffilmark{7}, L. Danaia\altaffilmark{4}, R. Edwards\altaffilmark{7} }


\altaffiltext{1}{Department of Physics \& Astronomy, Macquarie University, Sydney, Australia}
\altaffiltext{2}{Institute for Education Research, Edith Cowan University, Joondalup, WA, Australia}
\altaffiltext{3}{Denison College, Bathurst Campus, Australia}
\altaffiltext{4}{Charles Sturt University, Bathurst, Australia}
\altaffiltext{5}{UNIFEI, Instituto de F\'isica e Q\'imica, Universidade Federal de Itajub\'{a}, MG, Brazil}
\altaffiltext{6}{Instituto de Astronomia, Geof\'isica e Ci\^encias Atmosf\'ericas, Universidade de S\~ao Paulo, Cidade Universit\'aria,} 
S
\altaffiltext{7}{West Kildonan Collegiate, Winnipeg, Manitoba, Canada}\\


\begin{abstract}
Optical UBVRI photometric measurements using the Faulkes Telescope North were taken in early 2011 and combined with 2MASS JHK$_s$ and WISE infrared photometry as well as UCAC4 proper motion data in order to estimate the main parameters of the galactic open cluster NGC 2215 of which large uncertainty exists in the current literature. 
Fitting a King model we estimate a core radius of 1.12$'\pm$0.04$'$ (0.24$\pm$0.01pc) and a limiting radius of $4.3'\pm$0.5$'$ (0.94$\pm$0.11pc) for the cluster.
The results of isochrone fits indicates an age of $log(t)=8.85\pm0.10$ with a distance of $d=790\pm90$pc, a metallicity of $[Fe/H]=-0.40\pm0.10$ dex and a reddening of $E(B-V)=0.26\pm0.04$. 
A proportion of the work in this study was undertaken by Australian and Canadian upper secondary school students involved in the Space to Grow astronomy education project, and is the first scientific publication to have utilized our star cluster photometry curriculum materials.
\end{abstract}


\keywords{Methods: observational --- open clusters and associations : general --- open clusters and associations : individual (NGC 2215) --- Techniques: photometric}



\section{Introduction}
Open clusters have been used both for studies of stellar evolution and for dynamics and evolution of the Galactic disk. Compilations of fundamental parameters of these objects can be found in the catalogues of \cite{dia02} and WEBDA (Mermilliod 1988). However, of the 2174 open clusters cataloged only $\approx$400 clusters have been investigated with modern high quality CCD observations (Netopil et al. 2010), thus indicating the need in many cases to make further observations and analyses that allow a deeper, more precise and complete picture of stellar clusters within the Galaxy. This is especially true for NGC 2215, which is located in the third quadrant, a region that needs the largest number of results to improve the characterization of the Galaxy.

The open cluster NGC 2215 has a variety of diverging estimates for its distance, age, reddening and diameter over the last five decades with no previous metallicity estimates. These are outlined in Table 1. The first known study of open cluster NGC 2215 (Right Ascension 06$^h$20$^m$54$^s$, Declination -07$^{\circ}$17$'$ 42$''$,J2000.0 and galactic latitude 216.01417$^{\circ}$, galactic longitude -10.0896$^{\circ}$) was published by \cite{bec60} using data collected from photographic plates to produce the first color-magnitude diagram (CMD) containing 33 stars down to $V\approx$15.5 with $d=995$pc, $E(B-V)=0.10$ and $2.9$pc $(10')$ in diameter.

\begin{table}
\begin{center}
\caption{Previously Estimated Parameters for NGC 2215. \\
If later estimates appear to be quoted from earlier studies, these estimates have been left blank.\label{tbl-2}}
\begin{tabular}{cccccrrrrrrr}
\tableline\tableline
Reference&D&$E(B-V)$&Age&Diam\\
&(pc)&mag&log(t)&($'$)\\
\tableline
Becker (1960) & 995 &0.1 &  \\
Becker \& Feinhart (1971) & 1932 & 0.33  & & 18.8  \\
Maitzen et al. (1981) & & 0.244 & 7.6 \\
Pandey et al. (1989) & & & 8.55 \\
Frandsen \& Arentoft (1998) & 980 & 0.31 & 8.8 \\
Kharchenko et al. (2005) & 1298 & 0.3 & 8.43 & 33.6 \\
Loktin et al. (2001) & 1293 & 0.300 & 8.369 & 7 \\
Bukowiecki et al. (2011) & 1265 & 0.37 & 8.45 & 8.8 \\

\tableline
\end{tabular}
\end{center}
\end{table}

Few photometric follow up studies to the original conducted by \cite{bec60} are reported. One is indicated in \cite{bec71} as a personal communication although no actual paper has been able to be found. Their parameters were $d=1225$pc, $E(B-V)=0.33$ and a size of $3.6$pc ($10.1'$). \cite{mai81} report the \cite{bec60} distance but report  $E(B-V)=0.244$ using more sensitive \cite{stro66} photometry. They also estimate $log(t)=7.6$ from the (B-V) turnoff attributed to \cite{can70} who cites \cite{bec60} as the original source.

Later papers report discordant ages, distances and reddening. \cite{pan89} reports $log(t)=8.55$. \cite{per91} lists in his Table 1 data attributed to \cite{bec71} $d=1225$pc and $E(B-V)=0.33$. \cite{bec71}, however, report the distance as 1932pc and a diameter of 10pc (17.8$'$). \cite{fra98} report $log(t)=8.8$, $d=980pc$ and an $E(B-V)=0.31$. 

The Catalogue of Open Cluster Data (Kharchenko 2005) lists the data derived from ASCC-2.5 (Kharchenko et al. 2001) to this cluster as: ($E(B-V)=0.30$, $(m-M)=10.55$ (1293$pc$), $log(t)=8.43$ and diameter of the cluster of 6pc ($16.8'$), from only a small number (12) of brighter stars. Similarly, the data recorded in the \cite{dia02} catalogue of open clusters reports similar parameters, taken from \cite{lok01}  but with a visual estimate of the angular diameter of ($7'$), initially from \cite{lyn87}. A recent 2MASS IR study (Bukowiecki et al. 2011) estimates $d=1265pc$, $E(B-V)=0.37$, $log(t)=8.45$ and a diameter of $8.8'$.  These parameters appear to be inconsistent with those referred above. Thus, this paper attempts to clarify the disparate parameters attributed to this cluster.\\

%% In a manner similar to \objectname authors can provide links to dataset
%% hosted at participating data centers via the \dataset{} command.  The
%% second curly bracket argument is printed in the text while the first
%% parentheses argument serves as the valid data set identifier.  Large
%% lists of data set are best provided in a table (see Table 3 for an example).
%% Valid data set identifiers should be obtained from the data center that
%% is currently hosting the data.
%%
%% Note that AASTeX interprets everything between the curly braces in the 
%% macro as regular text, so any special characters, e.g. "#" or "_," must be 
%% preceded by a backslash. Otherwise, you will get a LaTeX error when you 
%% compile your manuscript.  Special characters do not 
%% need to be escaped in the optional, square-bracket argument.

\section{Observations}

\begin{figure}
\epsscale{1.25}
\plotone{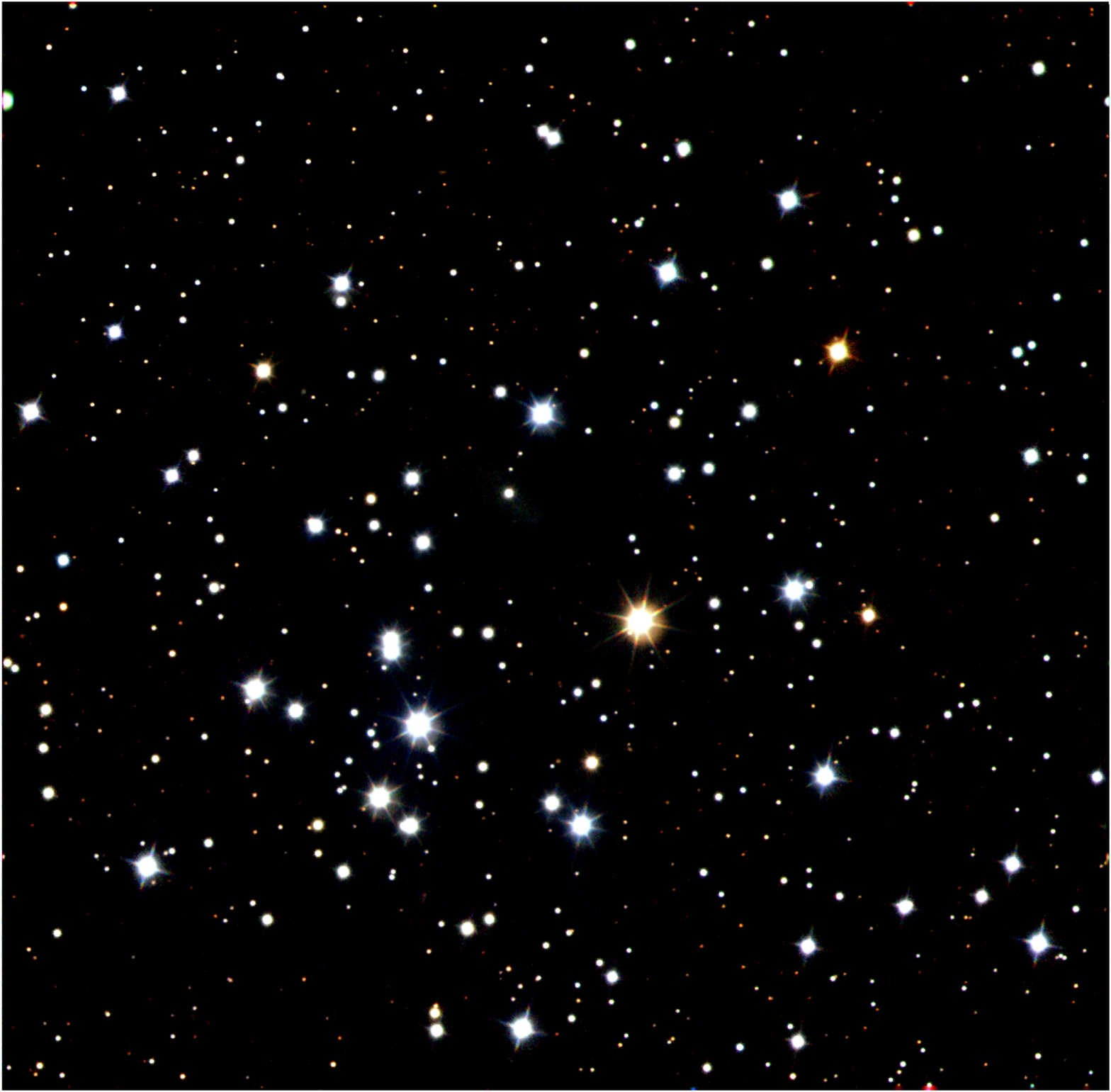}
\caption{BVR Color Image made from images used in this study. North is up, East is left. The field is roughly 9$'\times$9$'$.\label{fig1}}
\end{figure}

UBVRI observations of open cluster NGC 2215 were taken on January 27th 2011 using the Merope CCD Camera attached to the robotically controlled 2-metre Faulkes Telescope North at Haleakala, Hawaii operated by Las Cumbres Observatory Global Telescope (Brown et al. 2013). The pixel scale of the camera was $0.2785''/pixel$ in 2$\times$2 binning mode with a 4.7$'\times$4.7$'$ field of view. As the cluster itself was assumed to be larger than this field of view, five separate overlapping fields were taken with 3-5 short exposures per filter (U=200s, B=60s,V=40s,R=25s and I=15s) and 3 additional longer exposures (B=400s, V=300s, I=75s) of the central 4.7$'\times$4.7$'$. Bias and flat-field frames were taken and the science frames reduced at the telescope automatically prior to delivery to the observer. These images then had cosmic rays and bad pixels detected and removed using STARLINK (Disney \& Wallace 1982) internal routines as well as the L.A.Cosmic algorithm (Dokkum 2001). An accurate WCS in the ICRS for each image was obtained using astrometry.net software (Lang et al. 2010). A BVR mosaic of the full field of view is shown in Figure 1.  The typical seeing in all of these images was $\approx$1.2$''$.

Multiple observations of the SA98, RU149 and PG0918 Landolt standard fields (Landolt 1992, 2009, Clem \& Landolt 2013) surrounding the main observations were used to calibrate the images to the standard Johnson system using an ordinary linear squares regression fit. The calibration equations used are of the form:
\\
\\
\begin{eqnarray}
U = u + u_1 + u_2 × X + u_3 (U-B)\\
B = b + b_1 + b_2 × X + b_3 (B-V)\\
V = v + v_{1} + v_{2} × X + v_{3} (B-V)\\
R = r + r_1 + r_2 × X + r_3 (V-R)\\ 
I = i + i_1 + i_2 × X + i_3 (V-I)
\end{eqnarray}
where upper case letters represent the magnitudes and colors in the standard system and lower case letters were adopted for the instrumental magnitudes and X is the airmass. Observations made were only kept if there was a corresponding observation in a filter that facilitated a color correction. The range of airmass was quite short ($\approx$1.0 to $\approx$1.2) and the range of colors spanned from (B-V)$\approx$0 to $\approx$2.0. The multiple observations of the cluster itself are in the range of airmass 1.13 to 1.17.

The coefficient values are reported in Table 2, where the numbers in brackets refer to the error in the last figures of the provided coefficient. Figure 2 shows the differences between our observed photometry and the Clem \& Landolt (2013) catalogue values for, on average, 48 observations per filter. A photometric solution with $rms$ of $\approx$0.01 mags in UBVR and $\approx$0.02 in I were achieved. It is particularly notable that the U band has an uncommonly low color term. From four other observing nights using the same observational setup, the mean U band color term has been estimated to be $-0.033\pm0.012$ and the BVRI color terms are also similarly comparable to those obtained on this night.

\begin{figure}
   \centering
   \epsscale{1.2}
   \plotone{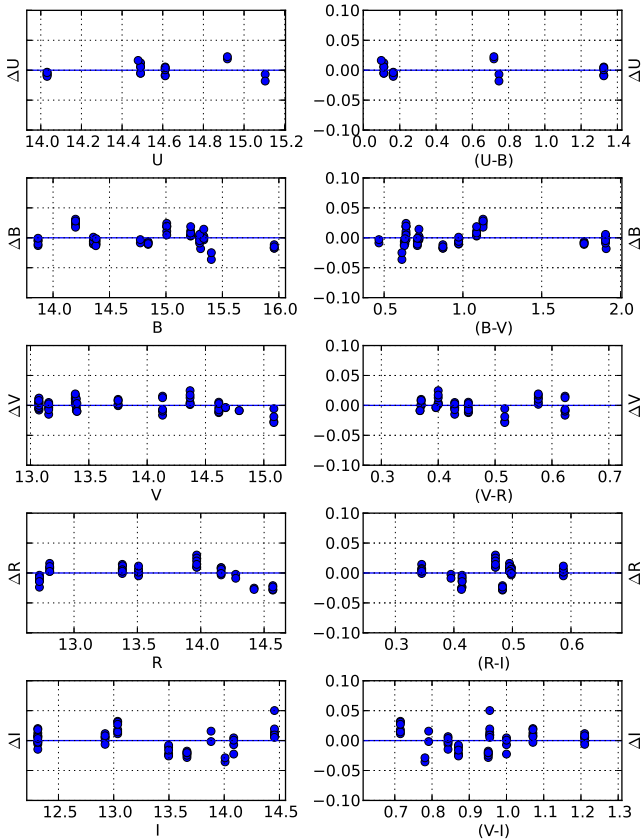}
   \caption{Residuals of the fit to the standard stars for the night. The \textit{rms} residuals of the transformations to the standard system are: $\Delta U = 0.011$; $\Delta B = 0.013$; $\Delta V = 0.010$; $\Delta R = 0.013$; $\Delta I = 0.018$.\label{fig2}}

   \label{correlation}%
    \end{figure}

%% In this section, we use  the \subsection command to set off
%% a subsection.  \footnote is used to insert a footnote to the text.


%% This section also includes several of the displayed math environments
%% mentioned in the Author Guide.

\begin{table}
%\tabcolsep 0.2truecm
\caption {Coefficients of the calibration equations.}
\centering
\begin{tabular}{cccc}
\hline
Zeropoint & Extinction & Color Term & $rms$\\
\hline
$u_1 = 22.119(43)$ & $u_2 =  -0.429(39)$ & $u_3 = -0.006(03)$ & $ 0.011$\\
$b_1 = 23.565(32)$ & $b_2 =  -0.190(29)$ & $b_3 = +0.046(03)$ & $0.013$\\
$v_{1} = 23.490(25)$ & $v_{2} =  -0.114(22)$ & $v_{3} = -0.084(02)$ & $0.010$\\
$r_1 = 23.391(42)$ & $r_2 =  -0.051(37)$ & $r_3 = -0.063(02)$ & $0.013$\\
$i_1 = 23.370(54)$ & $i_2 =  -0.023(47)$ & $i_3 = +0.064(01)$ & $0.018$\\
\hline

\hline
\end{tabular}
\end{table}
\section{Observational Parameters, Measurements and Results}

All astrometric and photometric measurements made in this study as well as proper motion data obtained from UCAC4 (Zacharias et al. 2013) and photometric data from 2MASS (Skrutskie et al. 2006) and WISE (Wright et al. 2010) are given in an online data file, with the format as shown in Table 3 contained within the Appendix. 

\subsection{Photometry} \label{bozomath}

Photometry of our images was undertaken via aperture photometry using Aperture Photometry Tool (APT) (Laher et al. (2012)). Aperture photometry using a 4 pixel radius (r$\approx$FWHM) aperture was performed using APT with aperture corrections for all measured stars. The sky was estimated for each star using the mode value per pixel for the local area of the image.

As there were multiple images taken of multiple overlapping fields, the number of measurements per star per filter range from 1 to 13 depending on their position in the field. These measurements were corrected for airmass and zeropoint then averaged together using the inverse of their estimated photometric error as weights to accommodate the different possible exposure times. This weighted mean magnitude corrected for airmass and zeropoint terms was then corrected for the color term.

\begin{figure}
\epsscale{1.15}
\plotone{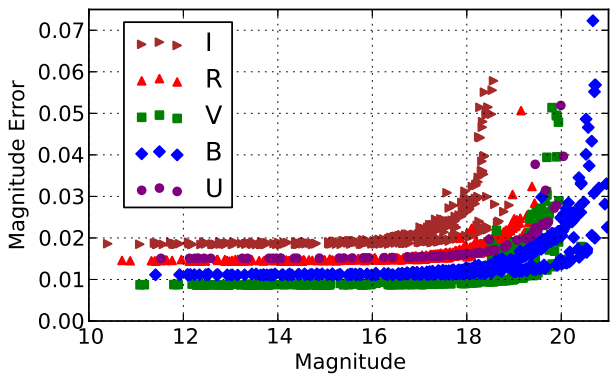}
\caption{Estimated photometric errors in the standard system in aperture photometry. \label{fig3}}
\end{figure}

Our instrumental photometric errors were combined with the errors propagated from the coefficients in the standard solution. The final errors are presented in Figure 3. 

\subsection{Comparison to Previous Photometry} \label{bozomath2}

\begin{figure}
\epsscale{1.15}
\plotone{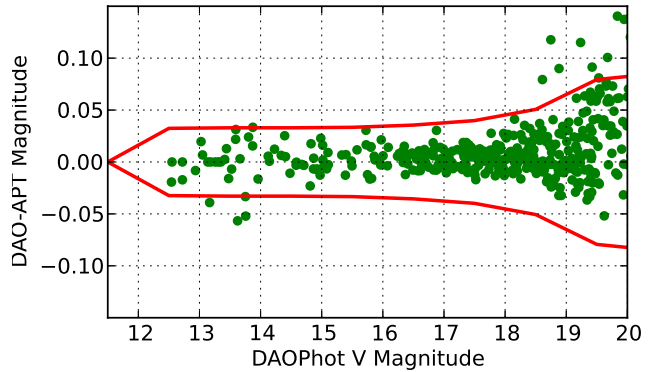}
\caption{Comparison between results from DAOPhot PSF and APT aperture photometry (APT). Dots are data points, line represents the 3$\sigma$ combined estimated total photometric error. \label{fig4}}
\end{figure}

As this is the first scientific use of APT to such depth that we are aware of, we compare our calibrated APT aperture photometry to calibrated PSF photometry using the latest version of DAOPhot (Stetson 1987) via the automated ALLPHOT (available from github.com/sfabbro/allphot) scripts. We find that very acceptable convergence between DAOPhot and APT. We present these results in Figure 4. 

We have compared our aperture photometry magnitudes to those available in roughly similar wavebands from all-sky surveys. We compare our B and V magnitudes to those available in the eighth data release of APASS (Henden et al. 2009). Our V magnitudes ($\Delta V=$0.012$\pm$0.033) and our B magnitudes ($\Delta B=$0.024$\pm$0.019) agree well with APASS magnitudes. Comparing our data to DENIS I (Epchtein et al. 1994, Deul et al. 1995) photometry, our results are not significantly (0.03$\pm$0.08) different.

A further night of observations were collected using the same  telescope and methodology as within this paper in March 2013 but only of the central 4.7$'\times$4.7$'$ of the field of view with only two science images per filter. The night was only borderline photometric with poorer ($\approx$2'') seeing, but had a larger airmass coverage ($\approx$1.0 to $\approx$1.5) and similarly high quality color coverage. Comparing the observations on the two nights shows that the 2011 observations used in this paper agreed with the later 2013 comparison images. The mean differences in each filter are: $\Delta U=-0.049\pm0.034,\Delta B=-0.012\pm0.051,\Delta V=-0.012\pm0.084,\Delta R=-0.034\pm0.040,\Delta I=-0.030\pm0.033$.

\subsection{Size of Cluster} \label{bozomath3}

To estimate the central co-ordinates and the size of the cluster, a King model (King 1962) was fitted to a Radial Density Profile (RDP) using both a traditional starcount method as well as a photometry based method. In the photometric method we essentially performed the radial count directly on the DSS image. For the starcount method we used the USNO-B1, 2MASS and WISE catalogues. For the photometric method we used three (blue, red and IR) DSS images.

We would have preferred to use our own CCD images and star counts as the source data, but as the cluster itself is on the order of the same angular size as the image, there was insufficient background to fit a King model.  We initially fit a King model by varying the core radius and peak density roughly by eye, then used least squares to find the best fit to the data. The specific King-like model used was that outlined by \cite{mac07} defined using $\rho(r)=f_{bg}+\frac{f_{0}}{1+\left(\frac{r}{r_{core}}\right)^{2}}$ and 
$r_{lim}=r_{core}\sqrt{\frac{f_{0}}{3\sigma_{bg}}-1}$, where $\sigma_{bg}$ is the background density, $f_0$ the central density of stars, and $r_{core}$ the core radius. 

\begin{figure}
\epsscale{1.15}
\plotone{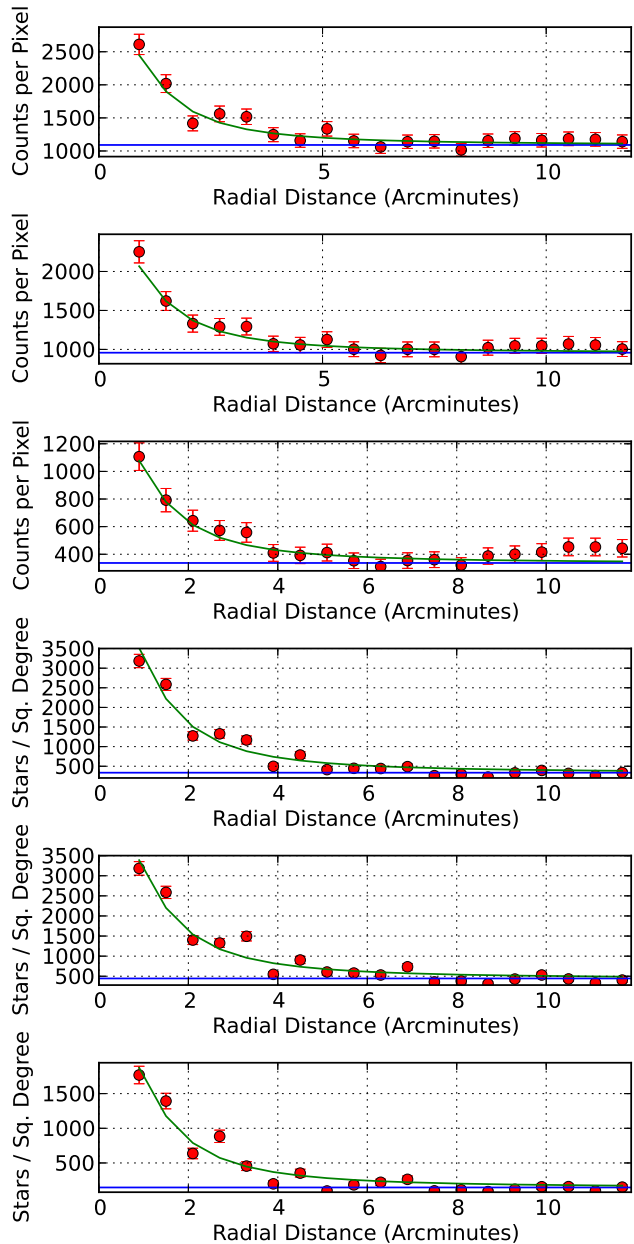}
\caption{The six radial density plots used to estimate the core and limiting radius of the cluster. From the top, the data sources are DSS Blue, DSS Red, DSS IR, WISE, 2MASS and USNO-B1\label{fig6}}
\end{figure}

Using the method outlined in \cite{mac07}, the central co-ordinates of the cluster were estimated to be Right Ascension 06$^h$20$^m$54$^s$, Declination -07$^{\circ}$17$'$42$''$. Individual King model fits to each RDP were made and are shown in Figure 5. From the mean values from these model fits, the core radius, $R_c=1.12'\pm0.04'$ and the limiting radius, $R_{lim}=4.3'\pm0.5'$ were determined leading to a concentration value, $c=R_c/R_{lim}$, of 0.26. The implied $8.6'\pm1.0'$ diameter is similar to those estimated qualitatively by \cite{dia02} and \cite{bec71}. \\
\\

\begin{figure}
\epsscale{1.0}
\plotone{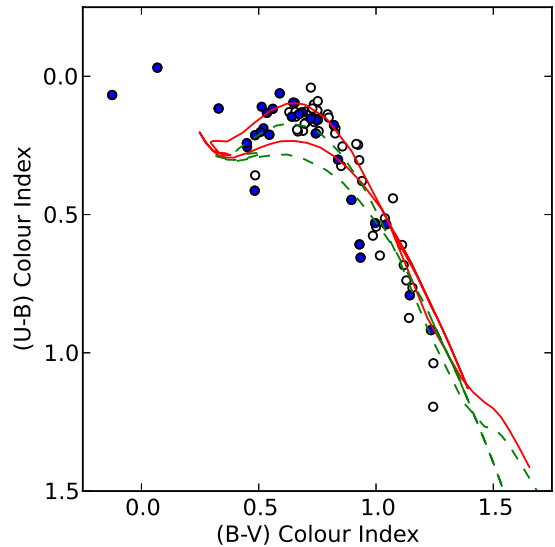}
\caption{Color-Color Diagram used for initial $E(B-V)$ and metalicity estimation. Dotted lines is the original solar metallicity isochrone, solid lines are corrected isochrones showing the metallicity estimation ($[Fe/H] = -0.4$ dex) via isochrone correction via the principle of the $(U-B)_{0.6}$ excess. White dots indicate the star has a low proper motion membership probability.  \label{fig7}}
\end{figure}

\subsection{Metallicity, Reddening and Extinction} \label{bozomath4}

(U-B) $vs$ (B-V) Color-Color diagrams were plotted initially against solar near-ZAMs ($log(t)=6.6$) isochrones (Girardi et al. 2002). The (U-B) $vs$ (B-V) diagram is well known to be very useful in estimating reddening but a less commonly utilised property of this diagram, is that it can also be used to estimate the metallicity of the cluster. We used the $\delta(U-B)_{0.6}$ ultraviolet excess method initially outlined by \cite{san69} and further refined by \cite{cam85} and \cite{kar06} by effectively exploiting the same principle by fitting the data using a grid of isochrones that vary in metallicity rather than focussing on a single deviation at a particular (B-V=0.6) color. 

We initially vary the E(B-V) using a solar metallicity isochrone to roughly fit our UBV data assuming a value for \textbf{$R_{v}$} of 3.1 (Winkler 1997) for which we obtained $E(B-V)\approx0.26$. At this stage we can increase the age of the isochrone to a rough lower age limit (log(t)$\approx8.8$) due to the lack of OB and early A type main sequence stars which results in a shortened isochrone with a slightly different shape. This is shown as the dotted line isochrone in Figure 6. We can then shift the metallicity to correct for the $\delta(U-B)_{0.6}$ ultraviolet excess, and in so doing be confident that we are finding a good estimate of the metallicity, as shown by the solid line in Figure 6. In this case, the best visual isochrone fit is Z=0.004, which translates into $[Fe/H]=-0.4$ $\pm$ 0.1 dex. Comparing the $\delta(U-B)_{0.6}$ of $\approx 0.1$ mag from this isochrone fit to the calibration of \cite{kar06}, we find an $[Fe/H]$ of $\approx -0.38$ dex, confirming our isochrone method is, essentially, very similar to the ultraviolet excess method. However, there is a subtle difference, in that the shape of the isochrone does subtly change at all colors with age, presumably leading to a variation in the ultraviolet excess and while this is a fairly small change, it would be non-zero.

Our overall reddening $E(B-V)=0.26\pm0.04$ is quite heavily constrained using this diagram. The error estimate is from visual inspection as the most extreme reddening that could be visually plausible for that particular metallicity. Most of the prior estimates of reddening from shallower broadband photometric data as shown in Table 1 are around $E(B-V)\approx0.3$, and \cite{sch98} dust maps imply an $E(B-V)$ of $0.372$, although this close to the Galactic plane this value can only be approximate at best and represents total extragalactic extinction rather than a typical within-Galaxy extinction. However the more reddening-sensitive Str\"{o}mgren photometry from \cite{mai81} is very close at $E(B-V)=0.244$, which gives confidence in our lower broadband reddening estimations. Also, this reddening seems quite typical, and in fact is roughly the mean value, for stellar clusters at these galactic co-ordinates (V\'{a}zquez et al. 2008).

\subsection{Mean Cluster Proper Motion, Membership Probability and Field Star Rejection} \label{bozomath5}

Astrometric reduction of our images were made using the astrometry.net software (Lang et al. 2010) while the positions were estimated from the pixel centroid outputs from APT. From these positions, the UCAC4 proper motions and 2MASS and WISE photometries were extracted via Vizier (Ochsenbein et al. 2000).

The optimum sampling radius of 4.5$'$ was determined using UCAC4 data following the recipe of \cite{san10}. This agrees satisfactorily with the limiting radius of the cluster presented previously. Following the method outlined in \cite{dia06} we applied the Zhao \& He (1990) statistical method to UCAC4 stars in the area using the central coordinates and optimum sampling radius previously mentioned.

Briefly, the method consisted in fitting the observed distribution of proper motions with two overlapping normal bivariate frequency functions, an elliptical one for the field stars and a circular one for the cluster stars, weighting the stellar proper motions with different errors. With the frequency function parameters we could determine the individual probability of the membership of each star in the cluster, as suggested by Zhao \& He (1990).

We obtained the mean proper motion for the cluster of  $\mu_{\alpha} cos \delta=+1.2\pm0.4$ mas/yr and  $\mu_{\delta}=-5.3\pm0.4$ mas/yr, the field proper motion of $\mu_{\alpha} cos \delta=-1.8\pm2.9$ mas/yr and  $\mu_{\delta}=-5.9\pm2.2$ mas/yr. These results compare well with previous estimates by \cite{dib02} from Tycho2 data of $\mu_{\alpha} cos \delta=+2.61\pm0.58$ mas/yr and $\mu_{\delta}=-5.60\pm0.58$ mas/yr. 

\begin{figure}
\epsscale{1.0}
\plotone{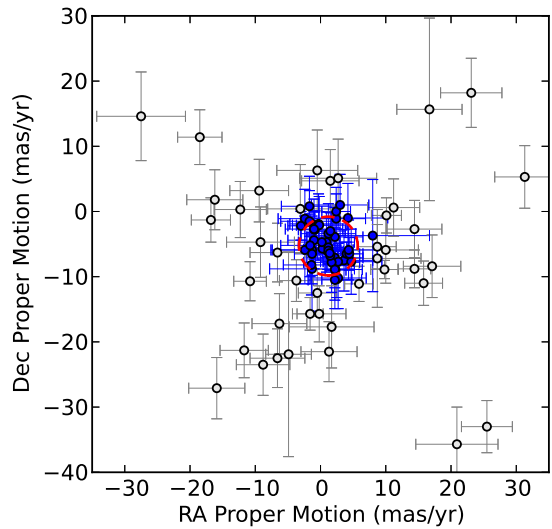}
\caption{Proper Motion Diagram of stars measured in UCAC4 catalogue within the region of radius of 4.5$'$ from the central coordinates of NGC2215. Filled circles have cluster probabilities above 0.61, empty circles have cluster probabilities below 0.61, the dashed circle represents the mean cluster proper motion 3$\sigma$ range. \label{fig8}}
\end{figure}

\begin{figure}
\epsscale{1.0}
\plotone{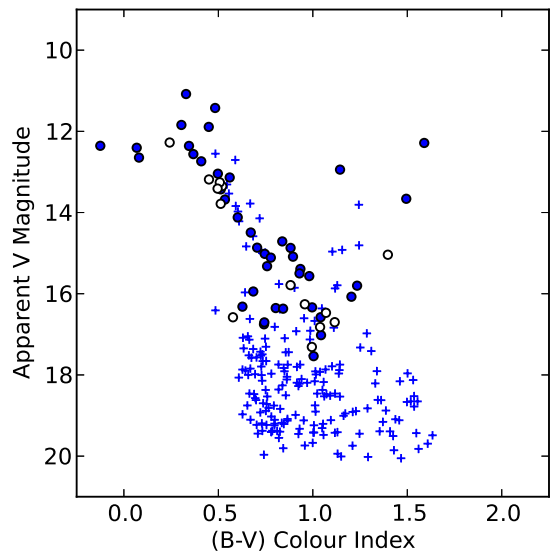}
\caption{Color Magnitude Diagram of all stars measured with photometric errors less than 0.1 mag and radial distance from the center of the cluster less than 4 arcminutes. Circular symbols represent stars rejected as non-members from the cluster, plus symbols represent stars with no proper motion data available, filled circles have cluster probabilities above 0.61.\label{fig9}}
\end{figure}

Figure 7 presents the vector proper motion diagram of the 105 UCAC4 stars in the cluster's region while also showing the field and cluster mean proper motions and standard deviations. In this work, we consider as kinematic members 51 stars with $P\geq61\%$.
There are seven higher proper motion stars within our field of view and measured photometrically that are outside the plotting bounds of Figure 7. A CMD showing the kinematic members is presented in Figure 8. Although there is still contamination of field stars due to limitation of the method and data, one can clearly see the signature of the cluster when considering only the kinematic member stars. This kinematic membership was used primarily to remove obvious non-members in order to achieve a more accurate visual fit.

\section{Combination of Optical Photometry with 2MASS and WISE}

In an endeavour to provide further constraints to our parameter estimates, we combined our optical photometry with near-IR data available from 2MASS and WISE all-sky surveys. After crossmatching the optical data with the infrared data, multiple CMDs across the entire optical/near-IR/mid-IR spectrum were used to visually fit isochrones using custom-designed software. It was found that comparing the optical CMDs to the infrared CMDs very heavily constrained the plausible values for stellar population parameters as even slight adjustments away from the optimal parameters led to large differences in quality of fit at opposite ends of the spectral range. 

The visual fit was performed simultaneously in the optical and IR considering as initial the pre-estimated values of E(B-V) and metallicity via the color-color diagram. To determine the fundamental parameters we adopted the extinction ratios provided by Cardelli et al. (1989), considering as usual $R_{V}= 3.1$.  We used isochrones of (Girardi et al. 2002) obtained from Padova database of stellar evolutionary tracks and isochrones.

\begin{sidewaysfigure*}
\centering
\epsscale{1.0}
\includegraphics[width = 20cm]{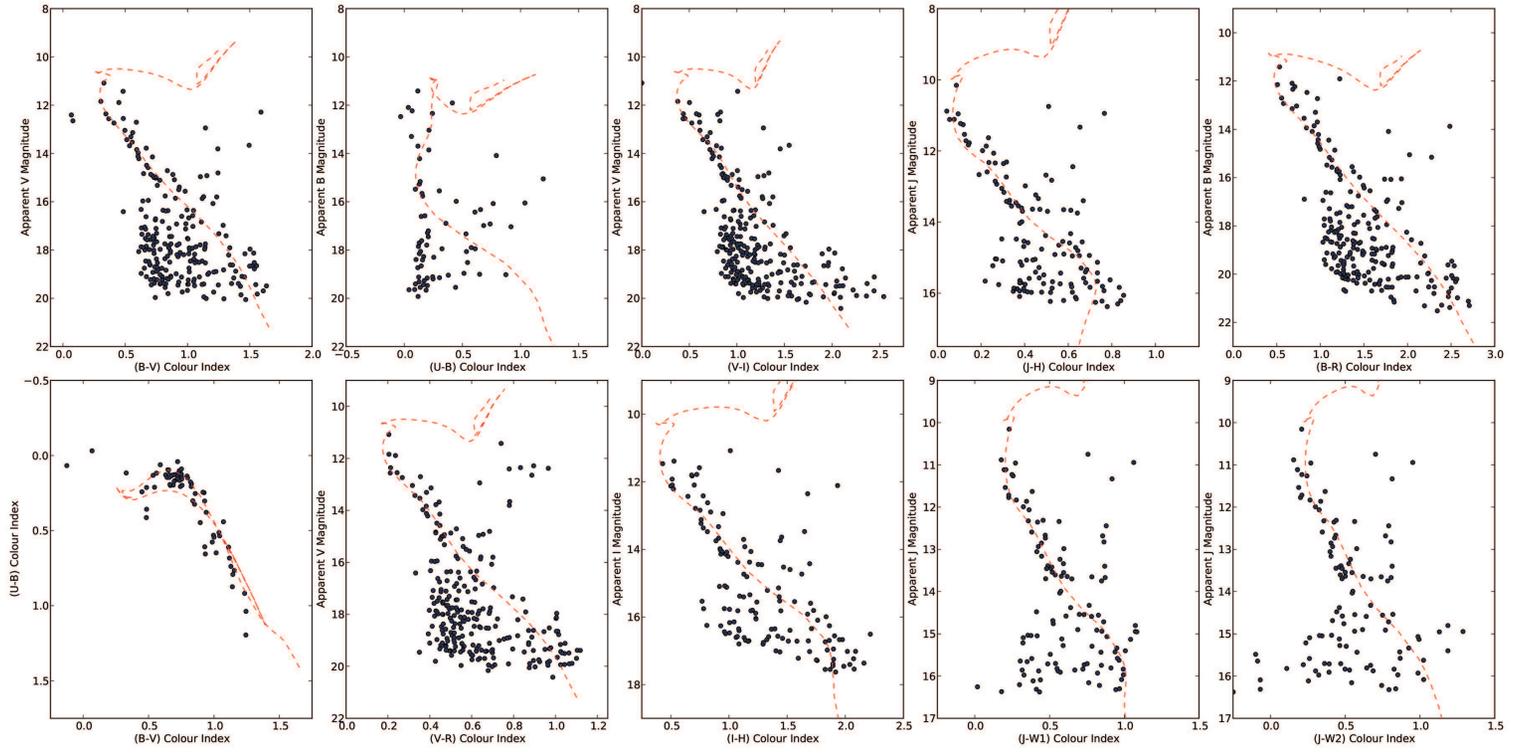}
\caption{Color Magnitude Diagrams of the NGC 2215. The UBVRI photometric measurements refes to our data, JH$K_{s}$ are data from 2MASS and W1 and W2 are data from WISE.
Overplotted are best fit isochrones from Padova models (Girardi et al. 2002) for distances 776pc, age of $log(t)=$ 8.8, the $E(B-V) = 0.26$ and  $[Fe/H] = -0.4$ dex.\label{fig20}}
\label{allplots}%
\end{sidewaysfigure*}

The CMDs with the final isochrone fits to the data are presented in Figure 9. The stellar population of the cluster is very heavily constrained as the bright end of the main sequence can be easily distinguished, with final parameters estimated to be $d=790\pm90$pc, $E(B-V)=0.26\pm0.04$, $log(t)=8.85\pm0.10$ and $Fe/H=-0.4\pm0.1$ dex. The uncertainties were
estimated to accommodate the values derived from the extreme visual fittings on simultaneous CMDs. A possible background population is apparent at ($log(t)\approx9.65$, $d\approx4$kpc, $E(B-V)\approx0.36$ and $Fe/H\approx-0.8$) which could be the population of the Perseus spiral arm. 

\section{Conclusion}

In this paper we have undertaken a UBVRI, 2MASS JHK$_s$ and WISE $W_1$/$W_2$ photometric data and UCAC4 proper motion to study the relatively neglected Galactic open cluster, NGC 2215. 
We have shown that the combination of optical and infrared data can be incredibly constraining in fitting stellar isochrones to observational data. While the distance parameter is relatively trivial, changes in metallicity, age or reddening parameters away from the optimal solution have fairly dramatic impacts on the quality of fit across the optical to infrared spectrum, heavily constraining the fit in a much stronger manner than using optical or infrared data alone. 
In a simultaneous visual fit we estimated the final parameters to be $d=790\pm90$pc, $E(B-V)=0.26\pm0.04$, $log(t)=8.85\pm0.10$ and $[Fe/H]=-0.4\pm0.1$ dex. 
This is the first estimate of the $[Fe/H]$ for the open cluster NGC 2215. Using the UCAC4 data the mean proper motion of NGC 2215 was estimated to be 
$\mu_{\alpha}cos\delta=+1.2\pm0.4$ mas/yr, $\mu_{\delta}=-5.3\pm0.4$ mas/yr. 

Applying the King model fit in the radial density profiles of multiple sources of data we estimate a core radius of 1.12$'\pm$0.04$'$ (0.24$\pm$0.01pc) and a limiting radius of $4.3'\pm$1.5$'$ (0.94$\pm$0.11pc) for the cluster. A large part of the initial scientific work within this project was undertaken by upper secondary school students involved in the Space to Grow astronomy education project (\cite{dan12}) in Australia and Canada.

\acknowledgments

We acknowledge the support of LCOGT.net whose provision of time on the Faulkes Telescopes has enabled this and other education/science crossover projects to take place. W. S. Dias acknowledges the S\~ao Paulo State agency FAPESP (fellowship 2013/01115-6). This research has made use of the VizieR catalogue access tool, CDS, Strasbourg, France. This research has made use of Aladin. In addition, this research has made use of the WEBDA database, operated at the Institute for Astronomy of the University of Vienna. This publication also makes use of data products from the Wide-field Infrared Survey Explorer, which is a joint project of the University of California, Los Angeles, and the Jet Propulsion Laboratory/California Institute of Technology, funded by the National Aeronautics and Space Administration. This publication makes use of data products from the Two Micron All Sky Survey, which is a joint project of the University of Massachusetts and the Infrared Processing and Analysis Center/California Institute of Technology, funded by the National Aeronautics and Space Administration and the National Science Foundation. Finally, this research made use of the cross-match service provided by CDS, Strasbourg. 


%% in parentheses, after the keyword, but they will not be verified.

{\it Facilities:} \facility{Faulkes Telescope North}.

\clearpage

\appendix

\section{Example of Sample Data}

\begin{sidewaystable*}
\begin{center}
\caption{Excerpt Sample of Online Data}
\begin{tabular}{cccccrrrrrrrcccccccccccc}
\tableline\tableline
&ID & RA & Dec  & U & err$_U$ & B & err$_B$ & V & err$_V$ & R & err$_R$ & I & err$_I$   \\

\tableline

&81&95.17041&-7.287&17.015&0.020&16.681&0.015&15.773&0.012&15.259&0.012&14.792&0.025\\
&82&95.17075&-7.266&nan&nan&19.610&0.016&18.955&0.012&18.533&0.015&18.115&0.025\\
&83&95.17084&-7.295&nan&nan&21.371&0.031&19.802&0.014&18.883&0.017&18.060&0.025\\
&84&95.17111&-7.320&nan&nan&18.775&0.017&17.915&0.014&17.376&0.014&16.910&0.026\\
&85&95.17118&-7.291&19.433&0.025&19.164&0.016&18.194&0.012&17.653&0.013&17.154&0.025\\
&86&95.17119&-7.318&nan&nan&19.177&0.018&18.165&0.014&17.543&0.014&16.983&0.026\\
&87&95.17122&-7.329&nan&nan&13.560&0.015&13.045&0.012&12.729&0.012&12.430&0.025\\
&88&95.17155&-7.278&16.765&0.019&16.644&0.015&15.952&0.012&15.527&0.012&15.096&0.025\\
&89&95.17253&-7.342&nan&nan&14.263&0.015&13.563&0.012&13.155&0.012&12.763&0.025\\
&90&95.17291&-7.284&17.630&0.019&16.929&0.015&15.780&0.012&15.141&0.012&14.527&0.025\\

\tableline\tableline
J & err$_J$ & H & err$_H$ & K$_s$ & err$_{K_s}$ & W1 & err$_{W1}$ & W2 & err$_{W2}$ & $\mu_{\alpha}$ & $\mu_{\delta}$& err$_{\mu_{\alpha}}$ & err$_{\mu_{\delta}}$ & Probability \\

\tableline
14.095&0.024&13.698&0.020&13.533&0.037&13.563&0.029&13.606&0.037&-0.2&-15.7&4.1&4.3&0.06\\
&&&&&&&&&&&&&&\\
16.782&0.128&16.216&0.149&15.491&0.200&16.288&0.080&17.199&&&&&&\\
16.088&0.075&15.653&0.111&15.417&&&&&&&&&&\\
16.368&0.093&15.730&0.108&16.314&&16.189&0.077&17.165&&&&&&\\
16.057&0.082&15.584&0.111&15.383&0.205&&&&&&&&&\\
11.992&0.024&11.785&0.020&11.716&0.019&11.667&0.025&11.691&0.024&-0.9&-2.8&2.5&2.9&0.91\\
14.474&0.063&14.179&0.046&14.039&0.072&&&&&&&&&\\
12.201&0.024&11.856&0.020&11.766&0.021&11.712&0.024&11.739&0.024&14.4&-8.8&2.5&2.9&0\\
13.668&0.029&13.056&0.020&12.901&0.030&12.796&0.026&12.861&0.029&&&&&\\

\tableline
\end{tabular}
\end{center}

From left to right, these values are our 1) ID\#, 2) Right Ascension in Degrees, 3) Declination in Degrees, 4 \& 5) U magnitude and error, 6 \& 7) B magnitude and error, 8 \& 9) V magnitude and error, 10 \& 11) R magnitude and error, 12 \& 13) I magnitude and error, 14 \& 15) 2MASS J magnitude and error, 16 \& 17) 2MASS H magnitude and error, 18 \& 19) 2MASS K magnitude and error, 20 \& 21) WISE W1 magnitude and error, 22 \& 23) WISE W2 magnitude and error, 24) Proper Motion in Right Ascension (mas/yr), 25) Proper Motion in Declination (mas/yr), 26) RA proper motion error, 27) Dec proper motion error, 28) Proper Motion Membership Probability

\end{sidewaystable*}

\clearpage

\end{document}